\documentclass[12pt]{article}
\usepackage{jheppub}
\usepackage[top=1in, bottom=1in, left=1in, right=1in]{geometry}
\usepackage[font=footnotesize]{caption}
\usepackage{array}
\usepackage{enumerate}
\usepackage{float}
\usepackage{subfig}
\usepackage{multicol}
\usepackage{multirow}
\usepackage{epsfig}
\usepackage{graphicx}
\usepackage{pict2e}
\usepackage{color}
\usepackage{amsmath,amsfonts,amssymb}
\usepackage{bbm}
\usepackage[utf8]{inputenc}
\usepackage[english]{babel}
\usepackage{xspace}

\newcommand{\be}{\begin{equation}}
\newcommand{\ee}{\end{equation}}
\newcommand{\bea}{\begin{eqnarray}}
\newcommand{\eea}{\end{eqnarray}}

\title{Draining the Swampland}

\begin{abstract}
{We discuss some implications of the recently suggested Swampland conjecture $\frac{|\nabla V|}{V}\gtrsim c\sim 1$, together with a previous one $\Delta \phi \lesssim 1$. We list some implications for particle phenomenology and the Early Universe.
The most intriguing implication of the conjecture could be a significant shift in allowed inflationary models, if not ruling out slow-roll (single field) inflation altogether. The tension of inflation and the conjecture does not only regard the amplitude of the tensor spectrum, but also its tilt, as $c\gtrsim 1$ implies both a yet unobserved tensor to scalar ratio, and an enhancement of the observed scalar power spectrum on large scales in discord with current data that favors a suppression on these scales. Scalar fields are abundant in theories of quantum gravity. Considering a second scalar field, its dynamics are dictated by the relation between its mass, $m$, and the Hubble parameter, $H$, at different epochs in the history of the Universe. This scalar field, a drainon, fulfills the conjecture draining up the swampland. For Inflation, this drainon requires a modest hierarchy compared to the inflaton. For the rest of the thermal history of the Universe, the drainon can be a coherently oscillating scalar field strengthening the case of Dark Matter candidates of that sort. 
}
\end{abstract}
\author{Ido Ben-Dayan}
\affiliation{
 Physics Department, Ariel University, Ariel 40700, Israel
}
\emailAdd{ido.bendayan@gmail.com}
\begin{document}

\maketitle
\section{Introduction}
Based on lingering debates about the consistency of de Sitter (dS) space and Quantum Gravity, it has been suggested that for any consistent theory of Quantum Gravity there exists an inequality \cite{Obied:2018sgi}:
\be
\textbf{C1:} \quad \frac{M_{pl}|\nabla V|}{V}\gtrsim c\sim 1 , \quad V>0.
\ee
The authors in \cite{Obied:2018sgi} give some supporting evidence for this claim, and in return it has been used to place bounds on the validity of inflationary and quintessence models \cite{Agrawal:2018own}. 
In the literature there are known counter examples to the conjecture, \cite{Kachru:2003aw, Brustein:2004xn,Balasubramanian:2005zx,Westphal:2006tn,BenDayan:2008dv,Rummel:2011cd}. However, there always seem to be lingering doubts about the validity of these constructions, for example, \cite{Danielsson:2018ztv,Sethi:2017phn}. Therefore, it makes sense to consider the implications of the conjecture on phenomenology, and whether the conjecture is true or may have to be revised.
 Let us stress that the argument is not limited to the specific point, but rather some domain in field space.
Therefore, the argument states that within a domain, $\Delta \phi$, the inequality holds. Furthermore, earlier conjectures based on the weak gravity conjecture and some explicit examples suggest 
\be
\textbf{C2:}\quad  \Delta \phi\lesssim M_{pl},
\ee
in field space \cite{Grimm:2018ohb}. Beyond  $\Delta \phi\sim M_{pl}$ additional light states are expected to appear in the spectrum of the theory invalidating the analysis.
Taken each criterion separately, phenomenological implications seem rather minimal. It may cast doubts on the validity of some inflationary models, but the literature is filled with constructions that fulfill one of the criterion or the other.

However, taking into account both criteria seems to severely constrain low energy effective field theories. Various suggestions and interpretations have been put forward  \cite{Denef:2018etk,Ghosh:2018fbx}. The interpretations involve several logical paths.
\begin{itemize}
\item Revising C1, for example, taking $c$ to be of the order of the slow-roll parameters $c\sim O(0.01)$. \item Revising $\Lambda CDM$ and/or Inflation, pointing towards quintessence as the cause of present acceleration and bouncing models rather than inflation as the mechanism that fits the known CMB data. 
\item Disproving the conjectures via explicit constructions
\item A combination of the above.
\end{itemize}
In this note, we analyze two approaches. 
First, we stick to \textit{strictly} single field analysis and discuss the implications of the criteria for Higgs Physics and Inflation.  Regarding the Higgs field, the criteria are in contradiction with the standard picture of spontaneous symmetry breaking and the electroweak phase transition. Regarding Inflation, we show that even C1 by itself, for any field range leads to interesting bounds on the tensor to scalar ratio $r$. The bound on $r$ is due to the correspondence between the slow-roll parameters and the CMB observables. 

Second, the spectrum of any realistic fundamental theory is \textit{never} a single field. The Standard Model fields are an immediate example. Even if single field inflation was realized in nature, fundamental theories and in particular string theory predict the existence of additional scalar fields, but their energy density and dynamics are negligible compared to the inflaton, which is why we can neglect them in deriving predictions of single field model or in analyzing inflationary dynamics. Specifically, the curvaton does not affect the inflationary dynamics, but its dynamics may have generated the observed CMB scalar spectrum \cite{Bartolo:2002vf}. Hence, to fulfill both criteria we have to take into account all fields.
We show how a second scalar field, a "drainon", allows to fulfill both criteria, hence draining the swampland. We apply the drainon idea for inflation and the thermal history of the Universe after inflation.
For the evolution of the Universe after inflation, the criteria plus simplicity point to the drainon being a coherently oscillating scalar field, strengthening the case for such Dark Matter models.
For inflation, the drainon is stuck at some point of the potential like a curvaton, but has no observable consequences.

\textbf{Note added:} While the preprint was being finalized \cite{Denef:2018etk} appeared which also showed how the Higgs potential is in contradiction with the dS swampland condition C1.



\section{Single Field Implications}
\subsection{Higgs, Spontaneous Symmetry Breaking and Particle Phenomenology}  
Taken at face value, it seems the criteria C1,C2 have far reaching consequences beyond what has been discussed, and might be a too restrictive requirement.
First, taken at face value, the bound precludes the existence of any extrema in the relevant field range $\Delta \phi$, since for both maxima and minima $\nabla V=0$ at finite, positive $V$.
Consider the simplest Mexican hat potential of spontaneous symmetry breaking:
\be
V(\phi)=\lambda(\phi^2-v^2)^2
\ee
where $v$ is the location of the minimum. For example, it is $v=246 GeV$ for the SM Higgs.
The bound  C1 works well if we take $\phi \rightarrow \infty$, (actually it works for $\phi >v$, thus fulfilling C2 as well), but it immediately fails for the maximum, where $\phi \rightarrow 0$.
Hence, the bound forbids such potentials as a good low energy effective description. 
Notice, that unlike inflation or quintessence that require $\Delta \phi \sim 1$, in the SM - the Mexican hat Higgs potential,  
the field difference between the two extrema has $\Delta \phi \sim 10^{-16}$.
If the bound is true, it means that such effective description is useless much earlier than $\Delta \phi \sim 1$.
Since the Mexican hat is a prototype of spontaneous symmetry breaking (SSB), the bound does not just put inflation in tension and rules out a Cosmological Constant. Rather, it also forbids a potential description of SSB, and phase transitions. Hence, to be consistent with C1, C2, we shall need a different description of Physics using different degrees of freedom.
To avoid such a consequence, one can refine the bound, by limiting it dS minima, i.e 
\be
\frac{|\nabla V|}{V}\geq c\sim 1\quad  or \quad \exists \lambda <0
\ee
where $\lambda$ is one of the eigenvalues of the Hessian, so extrema are possible, but no dS meta-stable minima.
In both cases, it will require the simple SM Higgs to be part of more comprehensive analysis that includes additional degrees of freedom 
 that are not at their minimum as long as their potential is positive. We will do so in the next section.

Since the bound  C1, is valid in some effective domain $\Delta \phi\sim 1$  according to C2, we can integrate it. The integral version of the bound yields:
\be
V(\phi)>V_0e^{c\phi} \quad or \quad 0\leq V(\phi)<V_0e^{-c \phi}
\ee
for some positive $V_0$.
So the only allowed positive potentials are very steep monotonic ones that grow or decay faster than exponentials. 

Considering strictly a single field, for instance the volume modulus, its real part involves decaying exponentials  $V\sim Ae^{-a T_R}$ based on gaugino condensation or instanton corrections \cite{BenDayan:2008dv}. The imaginary parts on the other hand enjoy a shift symmetry and appear with non-perturbative cosine potentials and a non-canonical kinetic term $\frac{(\partial T_I)^2}{T_R^2}$. To stabilze the volume modulus or have it play the role of the inflaton, we need some cancellation close to being a local extrema that the Universe can inflate off, or generate viable particle phenomenology. 
Prior to the discussed bound, some cancellation of these exponentials was used for model building. However, if the bound does not allow even such a delicate cancellation, it is hard to think of reasonable approximate low energy equilibrium phenomenology. The existence of the axionic partner of the volume modulus here does not help either, as it will contain an infinite number of isolated extrema that contribute zero to the gradient and we therefore have to make sure that the volume modulus has a steep enough exponential contribution, otherwise we may have an infinite number of isolated domains that are in the swampland, again making it hard to generate viable phenomenology.   Of course this approach is not realistic as string theory contains many moduli fields. By taking a multi-field approach, as in any realistic theory, one can circumvent the problematic bound C1. We shall demonstrate this in section $3$. 

\subsection{Inflation and C1}
Consider
\be
M_{pl} \frac{|\nabla V|}{V}\gtrsim c \sim 1
 \ee
 This is in one to one correspondence to the first slow-roll parameter $\epsilon$ even for multi-field models and for curved field spaces. Spelling the different terms explicitly
 \be
 \epsilon\equiv \frac{M_{pl}^2}{2} \frac{g^{ij}V_iV_j}{V^2}\gtrsim \frac{c^2}{2}
 \ee
 where $i,j$ run over all the relevant degrees of freedom, $g^{ij}$ is the inverse metric of the field space and a subscript denotes differentiation. If $c$ is really order unity, it seems slow-roll inflation is ruled out, and bouncing models are favored for two reasons. First, in bouncing models one needs "fast-roll", i.e. $\epsilon\sim c^2>1$, so this is fulfilled. Better yet, both for matter bounce and ekpyrotic scenarios, the potential is negative, so C1 is irrelevant all together. If so, then working alternatives require at least two fields, with various predictions including observable GW signal on CMB and/or Laser Interferometer scales \cite{Ben-Dayan:2016iks}.
 
 Focusing back on inflation, let us relax a bit C1 and C2, as these criteria are parametric, rather than strong absolute bounds. The reason is that C2 does not contradict a whole class of small field models and even within that class, observable $r$ is achievable \cite{BenDayan:2009kv}. So our working assumption is that C2 is easily fulfilled.
 Using the standard expressions:
\bea
n_s=1+2\eta-6\epsilon,\quad r=16\epsilon,\quad n_t=-2\epsilon
\eea
current data imposes $0.95\leq n_s\leq 0.97$ and $r<0.064$ \cite{Aghanim:2018eyx}.

We want inflation to be in accord with the data and C1, without any assumption on the distance travelled in field space. 
 Here we will see the crucial role of getting exact estimates on $c$.
 Considering a canonical single field, C1 states:
 \bea
 r&=&16\epsilon \gtrsim 8c^2 \Rightarrow c\lesssim \sqrt{0.008}\simeq 0.09 \label{rbound}\\
 n_s&=&1+2\eta-6\epsilon<1+2\eta-3c^2 \Rightarrow c\lesssim \sqrt{\frac{1-n_s+2\eta}{3}} \label{nsbound}\\
 n_t&=&-2\epsilon\lesssim-c^2
 \eea
 The second inequality \eqref{nsbound} gives a bound $\eta>-0.025$ that corresponds to $c=0$. Otherwise, the first inequality \eqref{rbound} is more stringent. Let us note, that C1 not only pushes $r$ to be as large as possible, it also makes $n_t$ more negative. Hence, it predicts enhancement of power on the largest scales, in discord with current measurements that favor some suppression of power, though at the level of $2-3\,\sigma$ \cite{Aghanim:2018eyx}.
 If $r$ was arbitrarily small, this enhancement wouldn't matter. But since C1 pushes $r$ to be as large as possible, it means this enhancement will be significant. If we limit the allowed enhancement to $10\%$, then we have another bound from $n_t$ giving $c<0.14$. In any case, both the tilt and the amplitude of GW in single field inflation disfavor $c\sim 1$ and push it down at least towards $c\sim 0.1$. 
 
We reach the following conclusion. If $c\sim 0.01$, then C1,C2 may rule out some models of inflation, but at least a whole class of small field models is valid. If $c\simeq 1$, then single field inflation is at odds with C1. Finally, there is the interesting regime of $c\sim 0.1$. If true, then we should expect a near future detection of $r$, and we need to address what models are in accord with the data, C1, and C2. 

Regarding model selection, even considering small field models, as we have explained in the context of Higgs physics, if C1 has any truth in it, then models inflating off an extrema such as $V(\phi)=\Lambda^4(1-a_2\phi^2+\cdots)$ are disfavored, since they will have $V_{\phi}\rightarrow 0$ as we approach the extrema, regardless of how small $c$ is. So even if we relax $c\sim 1$, it still favors small field models of inflection point type, as in \cite{BenDayan:2009kv}, and not inflating off an extremum. 

An interesting consequence of such limited inflation models, is the avoidance of eternal inflation and the multiverse, as that will require a region if field space with the spectrum of $P_s\sim V^3/M_{pl}^6V'^2\sim V c^2/M_{pl}^4\sim 1$ that is unlikely to exist in the region of $\Delta \phi \sim 1$ where the observed e-folds of inflation occurred \cite{Matsui:2018bsy}.
However, all these consequences are considering strictly single field, that is a toy model and not a realistic spectrum of particles. In the next section, when we introduce the drainon we shall see, that it allows us to push towards $c\sim 1$. Once the drainon field is taken into account, then at least all small field models are on the same footing again, and are viable models of inflation.

\section{The Drainon Field}
As we have explained, in any realistic theory, there are many fields in the spectrum. Thus, the criteria C1,C2 should be discussed while taking into account these additional fields. If we consider these additional fields, then the tension with SSB and phase transitions, quintessence and some inflationary models go away, as one field can be near an extrema, while the other is having an order unity gradient. Considering the field equation for a scalar field:
\be
\ddot{\phi}+3H\dot{\phi}+V'=0
\ee
If $H\gg m=V''$ then the field is stuck at some point in the potential, or if it is the inflaton it slowly rolls down the potential. However, if $m \gg H$ the field can be integrated out during inflation, or it coherently oscillates in other epochs. Depending on the properties of the potential the oscillating field can behave as various forms of matter. For a monomial potential, $V(\phi)=\lambda\frac{|\phi|^n}{n}$, the oscillating scalar behaves as energy density with the equation of state $w=(n-2)/(n+2)$, \cite{Cembranos:2015oya}. Hence, a massive free field behaves as dust, and can serve as a dark matter candidate. The field's oscillations do not have to be $\Delta \phi \sim 1$, they can be much less, then C2 proposes no problem to such models.

\subsection{The Post Inflationary Universe}
As the authors \cite{Obied:2018sgi} explicitly suggest, a massive scalar field is in accord C1, C2, 
\bea
V(\phi) =\frac{1}{2}m^2 \phi^2\cr
\Delta \phi<M_{pl} \Leftrightarrow \frac{M_{pl}V'}{V}=\frac{2M_{pl}}{\phi}>c\sim1  
\eea
taking a proper limit for $\phi=0$, and C1 and C2 are trivially fulfilled for $0<\phi\leq M_{pl}$. 
Hence, we can think of an oscillating scalar field. If we have a massive, approximately free field in our theory, then once the Hubble parameter drops below its mass $H<m$ the field will start to oscillate, and behave as dust with equation of state $w=0$. If our sole attempt is to fulfill C1 and C2, all we need is a scalar field with mass $m>T_{rh}^2/M_{pl}$, where $T_{rh}$ is the reheating temperature, and we are guaranteed C1,C2.
Better yet, since we are discussing properties of a would be quantum gravity theory, then our current best candidate, String Theory, actually predicts an Axiverse, i.e. that the particle spectrum is filled with axions on decades of mass all the way down to $m\sim 10^{-33} eV$ \cite{Arvanitaki:2009fg}. So at every decade from $T_{rh}$ we expect an axion with $m>H$ that will oscillate and therefore C1,C2 are fulfilled, at least from radiation domination until today even if there is no observable effect ever. Basically, this is the curvaton idea, where a light scalar field is stuck at some point on the potential during inflation, and later starts to oscillate and produces the observed CMB temperature fluctuations. Unlike the curvaton, here we do not need the drainon to produce the spectrum. 

 Since we are driven to consider an oscillating scalar field in our spectrum, for the sake for predictivity we can suggest the drainon as a DM candidate, for example, $m\simeq 10^{-22} eV$ and $f=10^{17} GeV$, with the 
axion potential
\be
V(\phi)=\Lambda^4\left(1-\cos\frac{\phi}{f}\right)
\ee
as recently analyzed in \cite{Rozner:2018ctk}.
If the axion DM actually acts as a drainon as well it has an interesting implication regarding the current acceleration and the CC.
Considering the Friedmann equation in $\Lambda$CDM, and neglecting radiation:
\be
\frac{H^2}{H_0^2}=\Omega_{m0}(1+z)^3+\Omega_{\Lambda0}
\ee
Since $\Omega_{\Lambda0}\sim \Omega_{m0}$, if $\Omega_{m0}$ is an oscillating scalar field, C1 and C2 are fulfilled, and current acceleration still can be explained as a CC.
Only if we consider the very far future when the energy density will be so small that even the lightest axions (perhaps as light as $m\sim 10^{-33}eV$) can be integrated out, and we will be strictly left with a CC, only then a CC is in contradiction with C1.

\subsection{Inflation and the Drainon}
Inflation is the paradigm with the biggest tension with C1 and C2.
Popular inflationary models typically have $\Delta \phi \gtrsim 1$, i.e. they violate C2. Worse, all simple single field models require $\sqrt{2\epsilon}=\frac{M_{pl}|V'|}{V}\ll1$ to be in accord with the CMB spectrum, while C1 says $\sqrt{2\epsilon}\equiv \frac{M_{pl}|V'|}{V}\gtrsim c\sim 1$. The fact that $\sqrt{2\epsilon}\sim c \ll 1$ is a requirement of small field models as well with $\Delta \phi \ll 1$.

A crucial question regarding inflation, is what fields participate in the gradient in C1. Since C1 and C2 are arguments regarding a low energy effective theory, there is a cut-off up to which the theory is trustable. Beyond that cut-off additional degrees of freedom or operators kick in, and have to be included in the analysis. In the context of Inflation, the lowest possible cut-off is given by the Hubble parameter, $H$. If the cut-off is of higher energy, it means more fields can contribute to the gradient and help fulfill C1. Hence, for a meaningful analysis, our working assumption will be that the cut-off is $H$ and the question whether low energy effective theories fulfill C1 or not, will include only degrees of freedom with mass lower than the Hubble parameter, $m<H$.
 
 The solution of an oscillating scalar field to fulfill C1 cannot work here as there are two options.
 Either $m\gg H$ and then we can integrate it out of the spectrum, or $m\ll H$, where we expect the field to be stuck or slowly roll. 
 If it is near the minimum, then the quantum fluctuations are larger than the classical behavior, and we cannot describe the behavior of the field as oscillating around the minimum \footnote{In such a case C1 is not well-posed, and stochastic inflation analysis is required that is beyond the scope of this note. But if the gradient is implying the magnitude of quantum fluctuations, then these quantum fluctuations are larger then the potential and C1 is fulfilled. Given minimal assumptions, this seems to be the behavior of the SM Higgs field during inflation \cite{Enqvist:2015sua}. }. If the field is at $\phi \gg H$ the semiclassical analysis is valid. However, in such a case it is difficult to have a field with negligible energy density, so it would not affect inflation on the one hand, and still actively contribute to the gradient of the potential to push $c$ towards unity on the other hand.
 
 Let us consider two scalar fields, the inflaton $\phi_1$ that is responsible for the observed power spectrum, and fulfills $\Delta \phi_1 \lesssim 1$, and a second field $\phi_2$, that its sole purpose is a drainon, such that $\frac{M_{pl}|\nabla V|}{V}\sim 1$.\footnote{Hybrid Inflation models \cite{Linde:1993cn} are an immediate example of two field models, but they immediately fail C1 and C2 for the following reason. Until the waterfall transition, one field is heavy and can therefore be integrated out. The outcome is single field inflation with large field excursion in $\Delta \phi_1 \gg M_{pl}$ in discord with C2 and with $M_{pl}V'/V\ll1$ in discord with C1. If the second field is light, then at its minimum it does not contribute to the gradient, and away from the minimum behaves as a curvaton. It could be the drainon provided it follows the analysis described in the text. We thank the anonymous referee for raising the question.} For simplicity, let us consider a massive free field as the drainon. We assume the energy density of the universe to be dominated by the inflaton, i.e. $V(\phi_1) \gg 1/2 m^2\phi_2^2$.
 Hence, during slow-roll:
 \bea
 \label {eq:sr1} 3M_{pl}^2H^2\simeq V(\phi_1)\\
  \label{eq:neg} V(\phi_1) \gg \frac{1}{2} m^2\phi_2^2\\
 \label{eq:hm} H\gg m\\
\label{eq:swamp} \frac{M_{pl}|\nabla V|}{V}=M_{pl}|\frac{V_{\phi_1}}{V}+\frac{m^2\phi_2}{V}|
  \eea
 Substituting \eqref{eq:sr1} and \eqref{eq:neg}  into \eqref{eq:swamp}, and denoting $\sqrt{2\epsilon}=M_{pl}|\frac{V_{\phi_1}}{V}|$, that shold give us standard slow-roll inflation, we get
 \be
 \frac{M_{pl}|\nabla V|}{V}=|\sqrt{2\epsilon}+\frac{m^2\phi_2}{3M_{pl}H^2}|\leq |\sqrt{2\epsilon}+\frac{m^2}{3H^2}|\sim c
\ee
 where in the inequality we have used $\phi_2\leq M_{pl}$ in accord with C2. So we see that $c\sim m^2/H^2\ll1$, that has to be less than unity. 
 
This limitation works well also with other forms of the potential like other powers or exponentials. Once we increase the derivative we also increase the potential, and then either we cannot neglect the energy density of the drainon, or it is not lighter than the Hubble parameter, or we cannot increase $c$ significantly beyond the standard slow-roll result, in discord with C1. 
The way out lies in quantum corrections. If the potential of the drainon is a logarithm, then its energy density increases logarithmically while its derivative increases with a power law. Such a potential can come from soft SUSY breaking and has been suggested as the running mass inflationary model \cite{Covi:1998mb}. Due to the non-polynomial nature of the potential, the model evaded some general effective field theory arguments about small and large field models \cite{BenDayan:2009kv}.
Consider the potential for the drainon:
\be
\label{eq:vlog}
V(\phi_2)=\Lambda^4\ln \left(\frac{\phi_2}{\phi_*}\right)
\ee
where $\Lambda$ and $\phi_*$ are some energy scales. For $\phi_2 \simeq \phi_*$ the energy density is completely negligible, and increases only logarithmically. 
For $\phi_*\lesssim M_{pl}$ we easily fulfill C2 for both the inflaton and the drainon throughout inflation. Let us repeat the previous exercise:
\bea
 \label {eq:sr1ln} 3M_{pl}^2H^2\simeq V(\phi_1)\\
  \label{eq:negln} V(\phi_1) \gg \Lambda^4\ln \left(\frac{\phi_2}{\phi_*}\right)\\
 \label{eq:hmln} H \gg \frac{\Lambda^2}{\phi_2}\\
\label{eq:swampln} \frac{M_{pl}|\nabla V|}{V}=M_{pl}|\frac{V_{\phi_1}}{V}+\frac{\Lambda^4/\phi_2}{V}|
  \eea
 Substituting \eqref{eq:sr1ln},\eqref{eq:negln} and \eqref{eq:hmln}  into \eqref{eq:swampln}, and denoting $\sqrt{2\epsilon}=M_{pl}\frac{|V_{\phi_1}|}{V}$, that should give us standard slow-roll inflation, we get
\be
 \frac{M_{pl}|\nabla V|}{V}=|\sqrt{2\epsilon}+\frac{\Lambda^4}{3M_{pl}H^2\phi_2}|<|\sqrt{2\epsilon}+\frac{\phi_2}{3M_{pl}}|\sim c
\ee
So, for $\phi_2\sim \phi_*\sim M_{pl}$ we get $c\sim 1$, the drainon has a completely negligible energy density so it does not affect the global evolution of the Universe and its mass is smaller than the Hubble parameter, so the field cannot be integrated out and it exists in the field spectrum. On the other hand, it comes close to saturate the bound of $ \frac{M_{pl}|\nabla V|}{V}\sim c\sim \mathcal{O}(1)$. Hence, we have shown why the drainon drains most of the swampland for both inflation and post-inflation evolution of the Universe. To come close to the bound $\Lambda^2\sim H \phi_2$, not being a very small energy scale. Nevertheless, after inflation $V_{\phi_2}$ will grow and as inflation ends C1 could be easily satisfied by the drainon or the inflaton depending on which of them will dominate the energy density of the Universe and reheat the universe. A simple minimum for a drainon field is if $V(\phi_2)=\Lambda^4\ln^2 \left(\frac{\phi_2}{\phi_*}\right)$. A stable Minkowski minimum is guaranteed at $\phi=\phi_*$, while the constraints on $H,\phi_2,\phi_*$ and $\Lambda$ do not change considerably. The dynamics are similar to \eqref{eq:vlog}. The logarithmic increase of energy density compared to a power law increase in the derivative allows to come close to saturating the bound without damaging inflation.

A realization of such a logarithmic potential in string theory is given in \cite{Hebecker:2011hk}, dubbed "fluxbrane inflation". By considering the relative position of two $D7$ branes  as the inflaton direction, the potential for the canonically normalized field is given by
\be
V(\phi_2)=\Lambda^4[1+\alpha \ln (\phi_2/\phi_*)]
\ee
providing a concrete realization of such type of potential in string theory. For our purposes $\phi_2$ does not have to play the role of the inflaton, but simply the drainon, which is easily fulfilled provided $V(\phi_2)\ll V(\phi_1)$ as we require.

Let us briefly discuss here a possible realization of the logarithmic potential in the supergravity approximation, not necessarily in the context of a two $D7$ branes.
Considering the following K\"ahler and superpotential
\bea
K=S \bar{S}+e^{T+\bar T}, \quad W=bST
\eea
then $S$ acts as a stabilizer field. Solving $D_TW=0$ forces $S=0$ as a result the lagrangian for the $T$ field reads
\be
 \mathcal{L}=e^{T+\bar T}\partial T\partial \bar T-b^2e^{e^{T+\bar T}}|T|^2
 \ee
 Canonically normalizing $u=e^T$ gives the lagrangian
 \be
  \mathcal{L}=\partial u\partial \bar u-b^2e^{u \bar{u}}|\ln u|^2
 \ee
For $u\ll1$,
\be
V\simeq b^2|\ln u|^2
\ee
which is qualitatively the same as \eqref{eq:vlog}. One can further write down $u=R e^{i\theta}$.
In that case $\theta=0$ is stabilized, and $R$ is canonically normalized leading to:
\be
V=b^2e^{R^2} \ln^2R \simeq_{R \ll 1} b^2 \ln^2R
\ee
which gives dynamics as \eqref{eq:vlog}.

\section{Summary}
The conjectures C1 and C2, and their predecessor the weak gravity conjecture, seem to pose an interesting challenge for Inflation, particle phenomenology and the Cosmological Constant. We have shown that the main challenge lies when assuming a model of a strictly single field, while considering a more realistic model allows to introduce a drainon that drains most of the swampland. Given an oscillating massive scalar, the problem of the present acceleration is pushed into the distant future, where $H<10^{-33} eV$, and the possibility of inflation is tenable given a drainon with a logarithmic potential.
The underlying motivation of the recent C1 and C2 conjectures is the difficulty in constructing a dS metastable vacua or inflationary model from the basic ingredients of a given string theory compactification. As such, having to accommodate a drainon field could be even more challenging. However, if model builders do mange to construct metastable dS or inflation in string theory explicitly, then the conjectures become irrelevant anyway and the drainon unnecessary. Given the weak evidence of the conjecture C1, the knowledge of the actual value of $c$ and the fact that a drainon seems to work rather well even for inflation, it seems premature to deviate from Inflation and $\Lambda$CDM that until now have been such a successful match to the data.


\begin{thebibliography}{99}
\bibitem{Obied:2018sgi} 
  G.~Obied, H.~Ooguri, L.~Spodyneiko and C.~Vafa,
  ``De Sitter Space and the Swampland,''
  arXiv:1806.08362 [hep-th].
  
\bibitem{Agrawal:2018own} 
  P.~Agrawal, G.~Obied, P.~J.~Steinhardt and C.~Vafa,
  ``On the Cosmological Implications of the String Swampland,''
  arXiv:1806.09718 [hep-th].
  
\bibitem{Kachru:2003aw} 
  S.~Kachru, R.~Kallosh, A.~D.~Linde and S.~P.~Trivedi,
  ``De Sitter vacua in string theory,''
  Phys.\ Rev.\ D {\bf 68}, 046005 (2003)
  [hep-th/0301240].
  
\bibitem{Brustein:2004xn} 
  R.~Brustein and S.~P.~de Alwis,
  ``Moduli potentials in string compactifications with fluxes: Mapping the discretuum,''
  Phys.\ Rev.\ D {\bf 69}, 126006 (2004)
  [hep-th/0402088].
  
\bibitem{Balasubramanian:2005zx} 
  V.~Balasubramanian, P.~Berglund, J.~P.~Conlon and F.~Quevedo,
  ``Systematics of moduli stabilisation in Calabi-Yau flux compactifications,''
  JHEP {\bf 0503}, 007 (2005)
  [hep-th/0502058].
  
\bibitem{Westphal:2006tn} 
  A.~Westphal,
  ``de Sitter string vacua from K\"{a}hler uplifting,''
  JHEP {\bf 0703}, 102 (2007)
  [hep-th/0611332].
  
\bibitem{BenDayan:2008dv} 
  I.~Ben-Dayan, R.~Brustein and S.~P.~de Alwis,
  ``Models of Modular Inflation and Their Phenomenological Consequences,''
  JCAP {\bf 0807}, 011 (2008)
  [arXiv:0802.3160 [hep-th]].
  I.~Ben-Dayan, S.~Jing, A.~Westphal and C.~Wieck,
  ``Accidental inflation from K\"{a}hler uplifting,''
  JCAP {\bf 1403}, 054 (2014)
  [arXiv:1309.0529 [hep-th]].
  I.~Ben-Dayan, F.~G.~Pedro and A.~Westphal,
  ``Towards Natural Inflation in String Theory,''
  Phys.\ Rev.\ D {\bf 92}, no. 2, 023515 (2015)
  [arXiv:1407.2562 [hep-th]].

\bibitem{Rummel:2011cd} 
  M.~Rummel and A.~Westphal,
  ``A sufficient condition for de Sitter vacua in type IIB string theory,''
  JHEP {\bf 1201}, 020 (2012)
  [arXiv:1107.2115 [hep-th]].
  
  
\bibitem{Danielsson:2018ztv} 
  U.~H.~Danielsson and T.~Van Riet,
  ``What if string theory has no de Sitter vacua?,''
  Int.\ J.\ Mod.\ Phys.\ D {\bf 27}, no. 12, 1830007 (2018)
  [arXiv:1804.01120 [hep-th]].
  
\bibitem{Sethi:2017phn} 
  S.~Sethi,
  ``Supersymmetry Breaking by Fluxes,''
  JHEP {\bf 1810}, 022 (2018)
  [arXiv:1709.03554 [hep-th]].
  
\bibitem{Grimm:2018ohb} 
  T.~W.~Grimm, E.~Palti and I.~Valenzuela,
  ``Infinite Distances in Field Space and Massless Towers of States,''
  arXiv:1802.08264 [hep-th].
  
\bibitem{Denef:2018etk} 
  F.~Denef, A.~Hebecker and T.~Wrase,
  ``The dS swampland conjecture and the Higgs potential,''
  arXiv:1807.06581 [hep-th].

\bibitem{Ghosh:2018fbx} 
  J.~K.~Ghosh, E.~Kiritsis, F.~Nitti and L.~T.~Witkowski,
  ``De Sitter and Anti-de Sitter branes in self-tuning models,''
  arXiv:1807.09794 [hep-th].
  D.~Andriot,
  ``New constraints on classical de Sitter: flirting with the swampland,''
  arXiv:1807.09698 [hep-th].
  C.~Roupec and T.~Wrase,
  ``de Sitter extrema and the swampland,''
  arXiv:1807.09538 [hep-th].
  A.~Ghalee,
  ``Condensation of a scalar field non-minimally coupled to gravity in a cosmological context,''
  arXiv:1807.08620 [gr-qc].
  S.~Paban and R.~Rosati,
  ``Inflation in Multi-field Modified DBM Potentials,''
  arXiv:1807.07654 [astro-ph.CO].
  R.~Brandenberger, L.~L.~Graef, G.~Marozzi and G.~P.~Vacca,
  ``Back-Reaction of Super-Hubble Cosmological Perturbations Beyond Perturbation Theory,''
  arXiv:1807.07494 [hep-th].
  E.~\'{O}.~Colg\'{a}in, M.~H.~P.~M.~van Putten and H.~Yavartanoo,
  ``$H_0$ tension and the de Sitter Swampland,''
  arXiv:1807.07451 [hep-th].
  M.~Dias, J.~Frazer, A.~Retolaza and A.~Westphal,
  ``Primordial Gravitational Waves and the Swampland,''
  arXiv:1807.06579 [hep-th].
  A.~Kehagias and A.~Riotto,
  ``A note on Inflation and the Swampland,''
  arXiv:1807.05445 [hep-th].
  J.~L.~Lehners,
  ``Small-Field and Scale-Free: Inflation and Ekpyrosis at their Extremes,''
  arXiv:1807.05240 [hep-th].
  S.~K.~Garg and C.~Krishnan,
  ``Bounds on Slow Roll and the de Sitter Swampland,''
  arXiv:1807.05193 [hep-th].
  A.~Ach\'{u}carro and G.~A.~Palma,
  ``The string swampland constraints require multi-field inflation,''
  arXiv:1807.04390 [hep-th].
  L.~Aalsma, M.~Tournoy, J.~P.~Van Der Schaar and B.~Vercnocke,
  ``A Supersymmetric Embedding of Anti-Brane Polarization,''
  arXiv:1807.03303 [hep-th].
  S.~Banerjee, U.~Danielsson, G.~Dibitetto, S.~Giri and M.~Schillo,
  ``Emergent de Sitter cosmology from decaying AdS,''
  arXiv:1807.01570 [hep-th].
  G.~Dvali and C.~Gomez,
  ``On Exclusion of Positive Cosmological Constant,''
  arXiv:1806.10877 [hep-th].
  D.~Andriot,
  ``On the de Sitter swampland criterion,''
  arXiv:1806.10999 [hep-th].
  
\bibitem{Bartolo:2002vf} 
  D.~H.~Lyth, C.~Ungarelli and D.~Wands,
  ``The Primordial density perturbation in the curvaton scenario,''
  Phys.\ Rev.\ D {\bf 67}, 023503 (2003)
  [astro-ph/0208055].
  N.~Bartolo and A.~R.~Liddle,
  ``The Simplest curvaton model,''
  Phys.\ Rev.\ D {\bf 65}, 121301 (2002)
  [astro-ph/0203076].
  A.~Mazumdar and J.~Rocher,
  ``Particle physics models of inflation and curvaton scenarios,''
  Phys.\ Rept.\  {\bf 497}, 85 (2011)
  doi:10.1016/j.physrep.2010.08.001
  [arXiv:1001.0993 [hep-ph]].
  
\bibitem{Ben-Dayan:2016iks} 
  I.~Ben-Dayan,
  ``Gravitational Waves in Bouncing Cosmologies from Gauge Field Production,''
  JCAP {\bf 1609}, no. 09, 017 (2016)
  [arXiv:1604.07899 [astro-ph.CO]].
  J.~L.~Lehners,
  ``Ekpyrotic Non-Gaussianity: A Review,''
  Adv.\ Astron.\  {\bf 2010}, 903907 (2010)
  [arXiv:1001.3125 [hep-th]].
  L.~A.~Boyle and A.~Buonanno,
  ``Relating gravitational wave constraints from primordial nucleosynthesis, pulsar timing, laser interferometers, and the CMB: Implications for the early Universe,''
  Phys.\ Rev.\ D {\bf 78}, 043531 (2008)
  [arXiv:0708.2279 [astro-ph]].
  M.~Gasperini,
  ``Observable gravitational waves in pre-big bang cosmology: an update,''
  JCAP {\bf 1612}, no. 12, 010 (2016)
  [arXiv:1606.07889 [gr-qc]].
  R.~H.~Brandenberger,
  ``String Gas Cosmology: Progress and Problems,''
  Class.\ Quant.\ Grav.\  {\bf 28}, 204005 (2011)
  [arXiv:1105.3247 [hep-th]].
  R.~Brandenberger and P.~Peter,
  ``Bouncing Cosmologies: Progress and Problems,''
  Found.\ Phys.\  {\bf 47}, no. 6, 797 (2017)
  [arXiv:1603.05834 [hep-th]].
  L.~A.~Boyle, P.~J.~Steinhardt and N.~Turok,
  ``The Cosmic gravitational wave background in a cyclic universe,''
  Phys.\ Rev.\ D {\bf 69}, 127302 (2004)
  [hep-th/0307170].
  E.~I.~Buchbinder, J.~Khoury and B.~A.~Ovrut,
  ``New Ekpyrotic cosmology,''
  Phys.\ Rev.\ D {\bf 76}, 123503 (2007)
  [hep-th/0702154].
  
\bibitem{BenDayan:2009kv} 
  I.~Ben-Dayan and R.~Brustein,
  ``Cosmic Microwave Background Observables of Small Field Models of Inflation,''
  JCAP {\bf 1009}, 007 (2010)
  [arXiv:0907.2384 [astro-ph.CO]].
  I.~Wolfson and R.~Brustein,
  ``Most probable small field inflationary potentials,''
  arXiv:1801.07057 [astro-ph.CO].
  I.~Wolfson and R.~Brustein,
  ``Small field models with gravitational wave signature supported by CMB data,''
  PLoS One {\bf 13}, 1 (2018)
  [arXiv:1607.03740 [astro-ph.CO]].
  S.~Hotchkiss, A.~Mazumdar and S.~Nadathur,
  ``Observable gravitational waves from inflation with small field excursions,''
  JCAP {\bf 1202}, 008 (2012)
  [arXiv:1110.5389 [astro-ph.CO]].
  
\bibitem{Aghanim:2018eyx} 
  N.~Aghanim {\it et al.} [Planck Collaboration],
  ``Planck 2018 results. VI. Cosmological parameters,''
  arXiv:1807.06209 [astro-ph.CO].
  Y.~Akrami {\it et al.} [Planck Collaboration],
  ``Planck 2018 results. X. Constraints on inflation,''
  arXiv:1807.06211 [astro-ph.CO].
  
\bibitem{Matsui:2018bsy} 
  H.~Matsui and F.~Takahashi,
  ``Eternal Inflation and Swampland Conjectures,''
  arXiv:1807.11938 [hep-th].

\bibitem{Cembranos:2015oya} 
  J.~A.~R.~Cembranos, A.~L.~Maroto and S.~J.~N\'{u}\~{n}ez Jare\~no,
  ``Cosmological perturbations in coherent oscillating scalar field models,''
  JHEP {\bf 1603}, 013 (2016)
  [arXiv:1509.08819 [astro-ph.CO]].
  
\bibitem{Arvanitaki:2009fg} 
  A.~Arvanitaki, S.~Dimopoulos, S.~Dubovsky, N.~Kaloper and J.~March-Russell,
  ``String Axiverse,''
  Phys.\ Rev.\ D {\bf 81}, 123530 (2010)
  [arXiv:0905.4720 [hep-th]].
  
\bibitem{Rozner:2018ctk} 
  M.~Rozner and V.~Desjacques,
  ``Backreaction of axion coherent oscillations,''
  Phys.\ Rev.\ D {\bf 98}, no. 2, 023530 (2018)
  [arXiv:1804.10417 [astro-ph.CO]].
  
\bibitem{Enqvist:2015sua} 
  K.~Enqvist, S.~Nurmi, S.~Rusak and D.~Weir,
  ``Lattice Calculation of the Decay of Primordial Higgs Condensate,''
  JCAP {\bf 1602}, no. 02, 057 (2016)
  [arXiv:1506.06895 [astro-ph.CO]].
  
\bibitem{Linde:1993cn} 
  A.~D.~Linde,
  ``Hybrid inflation,''
  Phys.\ Rev.\ D {\bf 49}, 748 (1994)
  [astro-ph/9307002].
  
\bibitem{Covi:1998mb} 
  L.~Covi and D.~H.~Lyth,
  ``Running mass models of inflation, and their observational constraints,''
  Phys.\ Rev.\ D {\bf 59}, 063515 (1999)
  [hep-ph/9809562].
  
\bibitem{Hebecker:2011hk} 
  A.~Hebecker, S.~C.~Kraus, D.~Lust, S.~Steinfurt and T.~Weigand,
  ``Fluxbrane Inflation,''
  Nucl.\ Phys.\ B {\bf 854}, 509 (2012)
  [arXiv:1104.5016 [hep-th]].
  A.~Hebecker, S.~C.~Kraus, M.~Kuntzler, D.~Lust and T.~Weigand,
  ``Fluxbranes: Moduli Stabilisation and Inflation,''
  JHEP {\bf 1301}, 095 (2013)
  doi:10.1007/JHEP01(2013)095
  [arXiv:1207.2766 [hep-th]].
  D.~Baumann and L.~McAllister,
  ``Inflation and String Theory,''
  arXiv:1404.2601 [hep-th].
  
  
\end{thebibliography}
\end{document}